\documentstyle[aps,12pt]{revtex}

\begin{document}
\draft

\title{Kinetic limit of $N$-body description \\ 
       of wave-particle self-consistent interaction}

\author{Marie-Christine Firpo\thanks{firpo@newsup.univ-mrs.fr}\ 
    and Yves Elskens\thanks{elskens@newsup.univ-mrs.fr}}

\address{Equipe turbulence plasma de l'UMR 6633 
         CNRS--Universit\'e de Provence\thanks{Address from October 1997: 
                service B22, case 321,
                Centre de Saint-J\'er\^ome, av. escadrille Normandie-Niemen,
                F-13397 Marseille Cedex 20}, \\
         IMT Ch\^ateau-Gombert, F-13451 Marseille cedex 20}


\maketitle
\vskip 0.5cm
\begin{abstract}
\parbox{14cm}

A system of $N$ particles $\xi ^{N}=(x_{1},v_{1},...,x_{N},v_{N})$
interacting self-consistently with $M$ waves $Z_{n}=A_{n}\exp (i\phi _{n})$
is considered. Given initial data $(Z^{N}(0),\xi ^{N}(0))$, it evolves according
to hamiltonian dynamics to $(Z^{N}(t),\xi ^{N}(t))$. 
In the limit $N\to\infty $, this generates 
a Vlasov-like kinetic equation for the distribution
function $f(x,v,t)$, abbreviated as $f(t)$, coupled to envelope equations
for the $Z_{n}$: initial data $(Z(0),f(0))$ evolve to $(Z(t),f(t))$. 
The solution $(Z,f)$ exists and is unique for any initial data with finite energy.
Moreover, for any time $T>0$, given a sequence of initial data with $N$
 particles distributed so that the particle distribution $f^{N}(0)\to f(0)$
  weakly and with $Z^{N}(0)\to Z(0)$ as $N\to
\infty$, the states generated by the hamiltonian dynamics at all times 
$0\leq t\leq T$ are such
that $(Z^{N}(t),f^{N}(t))$ converges weakly to $(Z(t),f(t))$.
\end{abstract}

\vskip 0.5cm 
\noindent PACS : 05.20.Dd (Kinetic theory)

\noindent 52.35.Fp (Plasma : electrostatic waves and oscillations)

\noindent 52.65.-y (Plasma simulation)

\noindent 52.25.Dg (Plasma kinetic equations)

\noindent Keywords : plasma, kinetic theory, wave-particle
interaction, mean-field limit

\clearpage 

\section{Introduction} Recent work on the dynamics of wave-particle
interaction has led to extensive use of $N$-body hamiltonian models in
parallel with the more traditional kinetic approach. The present paper aims
at discussing to what extent the two approaches agree in the limit $N \to
\infty$, where $N$-body dynamics formally reduces to kinetic theory. This is
a classical problem of statistical physics, notoriously unsolved for
particles interacting through short-range forces, where it amounts to
deriving the Boltzmann equation from the Liouville equation : systematic
rigorous derivations of the kinetic equation from BBGKY hierarchy are still
lacking -- notwithstanding the pioneering work of Lanford and King, limited
to short timescales \cite{King75,Lanford75}, and recent advances \cite
{Cercignani94,Piasecki97}. However, for long-range forces, and more
precisely for smooth enough mean-field interactions, the formal limit $N \to \infty$
commutes with the dynamics \cite{Neunzert84,Spohn91}. We show in this paper
how the mean-field methods apply also to wave-particle interactions.

A physical motivation for this work is that wave-particle interacting
systems are typical of plasmas and common to many physical phenomena. The
paradigm of such interactions is provided by the self-consistent hamiltonian 
$H_{sc}^{N,M}$ describing the evolution of $N$ particles and $M$ 
Langmuir waves \cite{Antoni93,Antoni96,Antoni97,Cary92,Doxas97,%
Escande90,Escande91,Guyomarch95,Mynick78,ONeil71,ONeil72,Tennyson94,Zekri93}. 
In particular this hamiltonian enables a mechanical approach of classical
plasma problems like Landau damping and beam-plasma instability, by treating
Langmuir waves as $M$ harmonic oscillators self-consistently coupled to $N$
quasiresonant beam particles. For simplicity we present our results in one
space dimension with periodic boundary conditions, which conforms to the
physical conditions considered in models of plasmas \cite
{Berk95,Fanelli96,Fried71}.

The basic characteristic of such models is that particles do not interact
directly with each other~: they only interact with the modes~;
symmetrically, the modes do not interact directly with each other~: they
only interact directly with the particles. Inasmuch the modes are spatial
Fourier components of some fields, these components are not localized
spatially~: this invites to describe the many-body limit $N \to \infty$ as a
mean-field limit and enables us to apply the techniques which succeed in the
case of particle-particle mean-field coupling.

The present work takes advantage of this observation to show that the
kinetic limit $N \to \infty$ and the time evolution over any time interval 
$[0,T]$ commute. Our result implies that numerical simulations with
increasing number of particles behave ever closer to the predictions of
kinetic theory (if one uses $N$ `large enough'...).

A preliminary form of our result was annouced in reference \cite{Firpo96b}.
In Sec. II we describe the model and its evolution equations. The main
results are stated in Sec. III. Sec. IV is devoted mainly to a finite $N$
estimate and a technical remark, preparing the proof presented in Sec. V.
The final section is devoted to the conclusion.

\section{Self-consistent hamiltonian and kinetic limit}

We consider a system of $N$ particles with respectively mass $m_{r}$, charge 
$q_{r}$, position $x_{r}$ and momentum $p_{r}$, interacting with $M$ waves
with respectively natural frequency $\omega _{j0}$, phase $\theta _{j}$ and
intensity $I_{j}$. The evolution of this system is described by the
hamiltonian 
\begin{equation}
H_{sc}^{N,M}=\sum_{r=1}^{N}{\frac{p_{r}^{2}}{2m_{r}}}+\sum_{j=1}^{M}\omega
_{j0}I_{j}-\varepsilon \sum_{r=1}^{N}\sum_{j=1}^{M}q_{r}k_{j}^{-1}\beta _{j}%
\sqrt{2I_{j}}\cos (k_{j}x_{r}-\theta _{j})  \label{Hsc}
\end{equation}
where the first term corresponds to free particles, the second term to free
waves (harmonic oscillators) and the third term to their coupling. The
coupling constants are expressed in such a way to ease the kinetic limit $%
N\to \infty $~: we shall keep the `wave susceptibilities' $\beta _{j}$
constant in this limit. A simple change of variables enables one to ensure
that all coefficients $\beta _{j}>0$, which is assumed in the following. 
The overall coupling factor $\varepsilon $ 
in the interaction term of (\ref{Hsc}) 
emphasizes our interest in the weak-coupling regime ($\varepsilon \ll 1$)
\cite{Antoni96,Escande96}.

Assuming periodic boundary conditions, the particles move on $({I\!\!R}/L)=S_L$ 
and the wavenumbers are quantized ($k_j=n_j 2 \pi /L$ for some integer $n_j$).
The phase space of this system is thus $(S_L \times {I\!\!R})^N \times {\cal Z}^M$
where ${\cal Z} = S_{2\pi} \times {I\!\!R}^{+}$ for each mode.

The natural scaling of our model in the limit $N\to \infty $ is easily
deduced from its equilibrium (Gibbs) thermodynamics \cite{Firpo96a}. Then
the energy $E=H_{sc}$ and the wave intensities $I_{j}$ are extensive (i.e. $%
{\rm O}(N)$), and the coupling constant scales as $\varepsilon ={\rm O}%
(N^{-1/2})$. The extensivity of wave intensities can easily be interpreted
as, in the physical regime of the model, we expect particles to be mostly
resonant with the waves, each such particle contributing then to wave
intensities by evolving in their potential well. This prompts us to
introduce intensive wave variables 
\begin{equation}
z_{j}=N^{-1/2}Z_{j}=N^{-1/2}\sqrt{2I_{j}}e^{-i\theta
_{j}}=|z_{j}|e^{-i\theta _{j}}  \label{defz}
\end{equation}
for which ${\cal Z}$ reduces to ${\raisebox{.6ex}{${\scriptscriptstyle /}$}%
\hspace{-0.43em}C}$, and renormalized coupling constants $\beta _{j}^{\prime
}=N^{1/2}\varepsilon \beta _{j}$.

The evolution equations of the hamiltonian (\ref{Hsc}) read 
\begin{eqnarray}
\dot x_r &=& p_r / m_r  \label{evolxN} \\
\dot p_r &=& {\frac{i }{2}} q_r \sum_{j=1}^M \beta^{\prime}_j ( z_j e^{i k_j
x_r} - z_j^{*} e^{- i k_j x_r} )  \label{evolpN} \\
\dot z_j &=& - i \omega_{j0} z_j + {\frac{i }{N}} \beta^{\prime}_j k_j^{-1}
\sum_{r=1}^N q_r e^{- i k_j x_r}  \label{evolzN}
\end{eqnarray}
To simplify calculations, we introduce non-canonical variables, namely 
particle velocities 
\begin{equation}
v_r = p_r / m_r  \label{defvr}
\end{equation}
and mode envelopes 
\begin{equation}
a_j = z_j e^{i \omega_{j0} t}  \label{defaj}
\end{equation}
bringing (\ref{evolxN}-\ref{evolpN}-\ref{evolzN}) to the form 
\begin{eqnarray}
\dot x_r &=& v_r  \label{evolxvN} \\
\dot v_r &=& {\frac{i q_r }{2 m_r}} \sum_{j=1}^M \beta^{\prime}_j ( a_j e^{i
k_j x_r - i \omega_{j0} t} - a_j^{*} e^{- i k_j x_r + i \omega_{j0} t} )
\label{evolvN} \\
\dot a_j &=& {\frac{i }{N}} \beta^{\prime}_j k_j^{-1} \sum_{r=1}^N q_r e^{-
i k_j x_r + i \omega_{j0} t}  \label{evolaN}
\end{eqnarray}

The usual space of kinetic theory is Boltzmann's $\mu $-space $\Lambda
=S_{L}\times {I\!\!R}$. The positions and velocities of the $N$ particles
determine a distribution $f$ on $\Lambda $ : 
\begin{equation}
f(x,v,t)={\frac{1}{N}}\sum_{r=1}^{N}\delta (x-x_{r}(t))\delta (v-v_{r}(t))
\label{distibution}
\end{equation}
which is normalized to unity ($\int_{\Lambda }f(x,v,t)dxdv=1$) irrespective
of the number $N$ of particles (for simplicity, we assume a single species~:
all $q_{r}=q>0$, $m_{r}=m$). The {\bf kinetic limit}, formally $N\to \infty $%
, corresponds to considering a sequence of $N$-particle distributions $f^{N}$
converging to a distribution $f^{\infty }$ in the weak sense for a natural
space of observables ${\cal D}$. Denote by ${\cal F}$ the space of positive
normalized distributions on $\Lambda $ with finite momentum and kinetic
energy, i.e. ${\cal F}\equiv \{f\in L^{1}(\Lambda ,dxdv): f \geq 0,
\int fdxdv=1,\int v^{2}fdxdv<\infty \}$ and define on ${\cal F}$
the bounded-Lipshitz distance 
\begin{equation}
d_{bL}(f,f^{\prime })\equiv \sup_{\phi \in {\cal D}}|\int_{\Lambda }\phi
fdxdv-\int_{\Lambda }\phi f^{\prime }dxdv|  \label{dbL}
\end{equation}
with the set of bounded, Lipschitz-continuous normalized observables 
\begin{equation}
{\cal D}\equiv \{\phi :\Lambda \to [0,1],|\phi (x,v)-\phi (x^{\prime
},v^{\prime })|\leq \Vert (x,v)-(x^{\prime },v^{\prime })\Vert \text{ }%
\forall (x,v),(x^{\prime },v^{\prime })\in \Lambda \}  \label{observables}
\end{equation}
Here $\Lambda $ is equipped with the distance $\Vert (x,v)-(x^{\prime
},v^{\prime })\Vert \equiv \alpha (|(x-x^{\prime }){\rm mod}L|+\tau |v-v^{\prime
}|)$, where $\alpha ^{-1}$ and $\tau $ are respectively convenient length
and time scales to be chosen below.

Then we consider the distance on ${\cal Z}^M$, 
\begin{equation}
\| a - a^{\prime}\| = \sum_{j=1}^M w_j |a_j - a^{\prime}_j|  \label{dista}
\end{equation}
where real positive coefficients $w_j$ will be chosen below in (\ref{wj}),
and $|a_j|$ is the modulus of the complex number $a_j$. Our distance on $%
{\cal F} \times {\cal Z}^M$ is just 
\begin{equation}
\| (f,a) - (f^{\prime},a^{\prime}) \| = d_{bL}(f,f^{\prime}) + \| a -
a^{\prime}\|  \label{distot}
\end{equation}

The kinetic evolution system of equations, dual to (\ref{evolxvN}-\ref
{evolvN}-\ref{evolaN}), is the system 
\begin{equation}
\partial_t f + v \partial_x f + {\frac{i q }{2 m}} \sum_{j=1}^M
\beta^{\prime}_j (a_j e^{i k_j x - i \omega_{j0} t} - a_j^{*} e^{- i k_j x +
i \omega_{j0} t}) \partial_v f = 0  \label{kinf}
\end{equation}
\begin{equation}
\dot a_j = i q \beta^{\prime}_j k_j^{-1} \int_\Lambda f(x,v,t) e^{-i k_j x +
i \omega_{j0} t} dx dv  \label{kina}
\end{equation}
This dynamics leaves ${\cal F} \times {\cal Z}^M$ invariant.

\section{Main results}

The self-consistent dynamics (\ref{evolxN}-\ref{evolpN}-\ref{evolzN})
preserves two constants of the motion, namely total energy $H$ and total
momentum $P = \sum_r p_r + \sum_j k_j I_j$. In the kinetic limit, we
consider the normalized constants $h=H/N$ and ${\sf p} = P/N$ : 
\begin{equation}
h(f, a) = \int_\Lambda {\frac{m v^2 }{2}}f dx dv + \sum_j \omega_{j0} {\frac{
|a_j|^2 }{2}} - \int_\Lambda \sum_j q k_j^{-1} \beta^{\prime}_j \Re (a_j
e^{i k_j x - i\omega_{j0} t} ) f dxdv  \label{defh}
\end{equation}
\begin{equation}
{\sf p}(f,a) = \int_\Lambda m v f dx dv + \sum_j k_j {\frac{ |a_j|^2 }{2}}
\label{defp}
\end{equation}
where $\Re$ denotes the real part. For any finite $N$ and $h$, the energy
surface $H_{sc}^{N,M} = N h$ in $\Lambda^N \times {\cal Z}^M$ is compact,
and the vector field (\ref{evolxN}-\ref{evolpN}-\ref{evolzN}) is continuous 
and bounded on it. This ensures that the dynamics generates a group 
for all initial conditions.

Moreover, the first variation of the dynamics (\ref{evolxvN}-\ref{evolvN}-%
\ref{evolaN}) generates a linear operator ${\cal M}=\partial (\dot{x}_{r},%
\dot{v}_{r},\dot{a}_{j})/\partial (x_{r},v_{r},a_{j})$, depending
continuously on $(x_{r},v_{r},a_{j})$. As the energy surface is compact for
any given $N$, ${\cal M}$ is bounded. With the specific form of $H_{sc}$, we
show that, with appropriate choice of the constants $w_{j}$~: 
\begin{equation}
\Vert {\cal M}\Vert \leq \tau ^{-1}+\gamma [a(t)]  \label{Mbound}
\end{equation}
where 
\begin{equation}
\tau =\Bigl({\frac{q^{2}}{m}}\sum_{j}{\beta _{j}^{\prime }}^{2}\Bigr)^{-1/3},
\label{deftau}
\end{equation}
\begin{equation}
\gamma [a(s)]=\sum_{j}{\frac{q\tau }{m}}\beta _{j}^{\prime }k_{j}|a_{j}(s)|
\label{dgamma}
\end{equation}
The positive function $\gamma [a(s)]$ is continuous on the energy surface,
on which it has an upper bound uniform with respect to $N$.

The kinetic limit, $N \to \infty$, admits a similar bound, ensuring the existence 
and uniqueness

{\bf Theorem} : Given initial data $(f_0,a(0)), (f^{\prime}_0,a^{\prime}(0))
\in {\cal F} \times {\cal Z}^M$, with $h_0 = h(f_0,a(0))$ and $h^{\prime}_0
= h(f^{\prime}_0,a^{\prime}(0))$, the kinetic evolution equations (\ref{kinf}%
-\ref{kina}) generate for all times $t \geq 0$ states $(f_t,a(t))$ and $%
(f^{\prime}_t,a^{\prime}(t))$ respectively from these data. Moreover, 
\begin{equation}
d_{bL}(f_t,f^{\prime}_t) + \|a(t)-a^{\prime}(t)\| \leq e^{C t}
(d_{bL}(f_0,f^{\prime}_0) + \|a(0)-a^{\prime}(0)\|)  \label{expothm}
\end{equation}
for some $C = C(h_0,h^{\prime}_0)< \infty$.

This theorem implies the

{\bf Corollary} : Given a distribution $f^\infty_0 \in {\cal F}$ and 
a sequence of finite-$N$ Dirac distributions $f^N_0 \in {\cal F}$
for particle initial data, such that $\lim_{N \to \infty}
d_{bL}(f^N_0,f^\infty_0) = 0$, given initial waves $a(0) \in {\cal Z}^M$, 
and given any time $T > 0$, consider for all $0 \leq t \leq T$ the 
resulting distributions $%
f^N_t $ and waves $a^N(t)$ generated by $H_{sc}^{N,M}$ and the kinetic
solution $(f^\infty_t,a^\infty(t))$. Then $\lim_{N \to \infty}
d_{bL}(f^N_t,f^\infty_t) = 0$ and $\lim_{N \to \infty} a^N(t) = a^\infty(t)$, 
uniformly on $[0;T]$.

In other words, the following diagram commutes for all $t > 0$~: 
\begin{equation}
\left. \matrix{ (f^N_0, a(0)) & {\buildrel
{(\ref{evolxvN}-\ref{evolvN}-\ref{evolaN})} \over \longrightarrow} & (f^N_t,
a^N(t)) \cr \qquad \quad \downarrow {N \to \infty} & & \qquad \quad
\downarrow {N \to \infty} \cr (f^\infty_0, a(0)) & {\buildrel
{(\ref{kinf}-\ref{kina})} \over \longrightarrow} & (f^\infty_t, a(t)) }%
\right.  \label{commutevol}
\end{equation}

\section{Preliminary remarks}

For given $N$ and finite energy $H$, the first variation ${\cal M}$ of the
dynamics (\ref{evolxvN})-(\ref{evolvN})-(\ref{evolaN}) has bounded norm
(with the $L_{1}$ distance) : 
\begin{equation}
\Vert (\delta \dot{x},\delta \dot{v},\delta \dot{a})\Vert _{1}\equiv
N^{-1}\sum_{r=1}^{N}\alpha (|\delta \dot{x}_{r}|+\tau |\delta \dot{v}%
_{r}|)+\sum_{j=1}^{M}w_{j}|\delta \dot{a}_{j}|  \label{maxM1}
\end{equation}
\[
=N^{-1}\sum_{r=1}^{N}\Vert (\delta \dot{x}_{r},\delta \dot{v}_{r})\Vert
+\sum_{j=1}^{M}w_{j}|\delta \dot{a}_{j}| 
\]
We readily find 
\begin{equation}
|\delta \dot{x}_{r}|=|\delta v_{r}|,  \label{dx}
\end{equation}
\begin{equation}
|\delta \dot{a}_{j}|=N^{-1}|\sum_{r=1}^{N}\beta _{j}^{\prime
}qe^{ik_{j}x_{r}-i\omega _{j0}t}\delta x_{r}|\leq N^{-1}\beta _{j}^{\prime
}q\sum_{r=1}^{N}|\delta x_{r}|  \label{da}
\end{equation}
and 
\begin{eqnarray}
|\delta \dot{v}_{r}| &=&\bigl| \sum_{j=1}^{M}{\frac{q\beta _{j}^{\prime }}{m}%
}\Re \bigl( e^{ik_{j}x_{r}-i\omega _{j0}t}(a_{j}k_{j}\delta x_{r}-i\delta
a_{j})\bigr) \bigr| \\
&\leq &\sum_{j=1}^{M}{\frac{q\beta _{j}^{\prime }}{m}}k_{j}|a_{j}|\cdot
|\delta x_{r}|+\sum_{j=1}^{M}{\frac{q\beta _{j}^{\prime }}{m}}\cdot |\delta
a_{j}|  \label{dv}
\end{eqnarray}
so that 
\begin{eqnarray}
\Vert (\delta \dot{x},\delta \dot{v},\delta \dot{a})\Vert _{1} &\leq
&N^{-1}\sum_{r=1}^{N}\alpha \tau ^{-1}(\tau \gamma [a(t)]\cdot |\delta
x_{r}|+\tau |\delta v_{r}|)  \nonumber \\
&&+\sum_{j=1}^{M}w_{j}N^{-1}\beta _{j}^{\prime }q\sum_{r=1}^{N}|\delta
x_{r}|+\alpha \tau \sum_{j=1}^{M}{\frac{q\beta _{j}^{\prime }}{m}}\cdot
|\delta a_{j}|  \label{maxM2}
\end{eqnarray}
with $\gamma [a(s)]$ defined by (\ref{dgamma}). 
The four causes for the divergence of trajectories 
in $\Lambda ^{N}\times {\cal Z}^{M}$ are 
saddle points (in $(x,v)$ plane) associated with maxima of the modes' potentials 
(the $|a_{j}|$ contribution to $\gamma [a(t)] $), 
velocity shear (the velocity term), 
the dependence of the modes source on the particle positions, 
and the dependence of the saddle points
themselves on the mode envelopes.

An appropriate choice of constants $\alpha $, $\tau $, $w_{j}$ keeps the
estimates as small as possible. Thus let 
\begin{equation}
w_{j}=\alpha w_{0}\beta _{j}^{\prime },  \label{wj}
\end{equation}
and solve 
\begin{equation}
\tau ^{-1}=w_{0}\sum_{j}q{\beta _{j}^{\prime }}^{2}={\frac{q\tau }{mw_{0}}}
\label{balance}
\end{equation}
This leads to the expression of $\tau $ announced in (\ref{deftau}) and to 
\begin{equation}
w_{0}=(qm)^{-\frac{1}{3}}\Bigl(\sum_{j}{\beta _{j}^{\prime }}^{2}\Bigr)^{-%
\frac{2}{3}},  \label{w0N}
\end{equation}
so that (\ref{maxM2}) reduces to 
\begin{equation}
\Vert (\delta \dot{x},\delta \dot{v},\delta \dot{a})\Vert \leq \tau
^{-1}\Vert (\delta x,\delta v,\delta a)\Vert +\gamma
[a(t)]N^{-1}\sum_{r}\alpha |\delta x_{r}|  \label{maxM3}
\end{equation}
which implies (\ref{Mbound}). Constant $\alpha $ remains arbitrary, as it
only determines the scale of the distances in $\Lambda $ and ${\cal Z}$, and
(\ref{Mbound}) is homogeneous (degree 1). Considering only the restricted dynamics on 
$\Lambda$, with $\delta a=0$, 
(\ref{maxM2}) straightforwardly leads to the continuity 
equation
\begin{equation}
  \Vert (\delta \dot{x},\delta \dot{v})\Vert 
  \leq 
  \gamma^{\prime}[a(t)] \Vert (\delta x,\delta v)\Vert
\label{gamprim}
\end{equation}
with
\begin{equation}
  \gamma^{\prime}[a(t)] = \max (\tau^{-1}, \gamma[a(t)]).
\label{defgamprim}
\end{equation}
Note that $\gamma^{\prime}[a(t)]$ is bounded uniformly in time, 
as the positive function $\gamma[a]$ is bounded above on the energy surface by
a function which does not grow faster than $h^{1/2}$ in the large energy
limit. More precisely, let $\lambda>0$ solve 
$\lambda^2 \sum_j \omega_{j0}^{-1} (q \beta^{\prime}_j k_j/m)^2 
  = 2 h + \sum_j \omega_{j0}^{-1} (q \beta^{\prime}_j / k_j)^2$. 
Then $\sum_j |\beta^{\prime}_j k_j a_j q/m| 
\leq (q^2 / m) 
\sum_j \omega_{j0}^{-1} {\beta^{\prime}_j}^2 (1 + k_j^2 \lambda / m)$. 
It should be noted 
once more that (\ref{gamprim}) reflects that 
the divergence rate in $(x,v)$ is controlled by velocity shear and by 
saddle points of the pendulum-like potential depending on wave amplitudes. 
The latter situation typically corresponds to a trapping regime for large 
enough wave intensities.

Finally, note the following

{\bf Proposition 1} : Let $Y:\Lambda \to \Lambda $ be a Lipschitz mapping
with constant $L\geq 1$ on $\Lambda $, and $\mu ,\nu \in {\cal F}$. Then : 
\begin{equation}
  \sup_{\phi \in {\cal D}}|\int_{\Lambda }\phi \circ Y d(\mu -\nu )|
  \leq
  L d_{bL}(\mu ,\nu )  
\label{pro1}
\end{equation}

{\bf Proof} : Clearly $L^{-1} \phi \circ Y \in {\cal D}$ 
for any $\phi \in {\cal D}$. 
Hence $\sup_{\phi \in {\cal D}} 
       | \int_\Lambda L^{-1} \phi \circ Y d(\mu - \nu) | 
       \leq d_{bL} (\mu, \nu)$ .

\section{Proof of the main result}

The proof of theorem 1 uses the fact that the two types of degrees of
freedom have no `self'-interaction. Indeed the motion of particle $r$ is
completely determined by its initial position and velocity and by the modes
history, i.e. the data of the modes $a_{j}(.)$ over a time interval $[s,t]$ 
defines the vector field $G$ so that: 
\begin{equation}
  {\frac{d}{dt}} (x_{r}(t),v_{r}(t)) = G[a(t)] (x_{r}(t),v_{r}(t))  
\label{Gxv}
\end{equation}
This vector field is Lipschitz-continuous on $\Lambda $ according to
 (\ref{gamprim}) and subsequent remarks. Thus Cauchy-Lipschitz theorem ensures
the existence and unicity of the flow $T$:
\begin{equation}
  (x_{r}(t),v_{r}(t))=T_{t,s}[a(.)](x_{r}(s),v_{r}(s))  
\label{Txv}
\end{equation}
By duality the measure $\mu _{s}$ on $\Lambda$ is transported by the flow to 
\begin{equation}
  \mu _{t}=\mu _{s}\circ T_{s,t}[a(.)]  
\label{defT}
\end{equation}
Similarly, the evolution of mode $j$ is also completely determined by its
initial data $a_{j}(s)$ and by the history of the measure on particle phase
space $\Lambda $ in the right hand side of (\ref{kina}) which defines a flow 
$S$ by 
\begin{equation}
   a_{j}(t) = S_{t,s}[\mu _{.}]a_{j}(s)  
\label{defS}
\end{equation}
Solving kinetic equations (\ref{kinf})-(\ref{kina}) with initial data $(\mu _{0},a(0))$
amounts to finding a fixed point of the coupled system (\ref{defT})-(\ref{defS}) in
the space $({\cal F}\times {\cal Z}^{M})^{{I\!\!R}}$. Our strategy now
follows that of Neunzert \cite{Neunzert84} and Spohn \cite{Spohn91}, who
considered direct particle-particle interaction of mean-field type.

Thus consider two solutions $(f_{.}, a(.))$ and $(f^{\prime}_{.},
a^{\prime}(.))$ of (\ref{kinf})-(\ref{kina}). To shorten notations, write $%
b=a^{\prime}$ and denote by $d\mu = f dx dv$ and $d\nu = f^{\prime}dx dv$
the corresponding measures. Their distance at time $t$ satisfies 
\begin{equation}
\| (\mu_{t}, a(t)) - (\nu_{t}, b(t)) \| = d_{bL}(\mu_0 \circ T_{0,t}[a(.)] ,
\nu_0 \circ T_{0,t}[b(.)] ) + \| S_{t,0} [\mu_{.}] a(0) - S_{t,0} [\nu_{.}]
b(0) \|  \label{equ1}
\end{equation}
where 
\begin{equation}
\| S_{t,0} [\mu_{.}] a(0) - S_{t,0} [\nu_{.}] b(0) \| \leq d_1(t) + d_2(t)
\label{ineq1}
\end{equation}
\begin{equation}
d_{bL}(\mu_0 \circ T_{0,t}[a(.)] , \nu_0 \circ T_{0,t}[b(.)] ) \leq d_3(t) +
d_4(t)  \label{ineq2}
\end{equation}
and 
\begin{equation}
d_1(t) = \| S_{t,0} [\mu_{.}] a(0) - S_{t,0} [\mu_{.}] b(0) \|  \label{defd1}
\end{equation}
\begin{equation}
d_2(t) = \| S_{t,0} [\mu_{.}] b(0) - S_{t,0} [\nu_{.}] b(0) \|  \label{defd2}
\end{equation}
\begin{equation}
d_3(t) = d_{bL} ( \mu_0 \circ T_{0,t}[a(.)] , \nu_0 \circ T_{0,t}[a(.)] )
\label{defd3}
\end{equation}
\begin{equation}
d_4(t) = d_{bL} ( \nu_0 \circ T_{0,t}[a(.)] , \nu_0 \circ T_{0,t}[b(.)] )
\label{defd4}
\end{equation}

Straightforward integration of (\ref{kina}) shows that 
\begin{equation}
  d_1(t) = d_1(0) = \| a(0) - b(0) \|  
\label{vald1}
\end{equation}
because the flow $S[\mu_{.}]$ is just a translation in ${\cal Z}^M$.

To estimate $d_2$ we integrate (\ref{kina}) with the right hand sides given
by $\mu_{.}$ and $\nu_{.}$ : 
\begin{eqnarray}
  d_2(t) &=& 
  \sum_{j=1}^M w_j q \beta^{\prime}_j k_j^{-1} 
  \Bigl| \int_0^t \int_\Lambda e^{-i k_j x + i \omega_{j0} s} 
                                    d(\mu_s - \nu_s) ds \Bigr| \\
  &=& 
  2 \sum_{j=1}^M w_j q \beta^{\prime}_j k_j^{-1} 
  \Bigl| \int_0^t \int_\Lambda 
        { {1 + i + e^{-i k_j x + i \omega_{j0} s} } \over {2}} 
        d(\mu_s - \nu_s) ds \Bigr| \\
  &\leq& 
  \sqrt{2} \tau ^{-1} \int_0^t d_{bL}(\mu_s,\nu_s) ds  
\label{vald21}
\end{eqnarray} 
In (\ref{vald21}) the inequality uses the fact that 
$\alpha (1 + \cos (k_j x - \theta))/k_j  \in {\cal D}$ and 
$\alpha (1 + \sin (k_j x - \theta))/k_j  \in {\cal D}$ 
for any real $\theta$, provided that $2 \alpha \leq \min k_j$.

Estimating $d_3$ is also straightforward, as proposition 1 implies 
\begin{equation}
d_3(t) \leq d_{31}(t) d_{bL}(\mu_0 , \nu_0)  \label{vald31}
\end{equation}
provided that $d_{31}(t)$ is a Lipschitz constant for $T_{t,0}[a(.)]$. Now, $%
\forall (x,v), (x^{\prime},v^{\prime}) \in \Lambda$, 
\begin{eqnarray}
& \| T_{t,0}[a(.)] (x,v) - T_{t,0}[a(.)] (x^{\prime},v^{\prime}) \| 
\nonumber \\
&\leq \| (x,v) - (x^{\prime},v^{\prime}) \| + \int_0^t \| G[a(s)] T_{s,0}[a]
(x,v) - G[a(s)] T_{s,0}[a] (x^{\prime},v^{\prime}) \| ds \\
&\leq \| (x,v) - (x^{\prime},v^{\prime}) \| + \int_0^t \gamma^{\prime}[a(s)]
\| T_{s,0}[a] (x,v) - T_{s,0}[a] (x^{\prime},v^{\prime}) \| ds
\label{vald32}
\end{eqnarray}
Hence, $d_{31}(t) \leq 1 + \int_0^t \gamma^{\prime}[a(s)] d_{31}(s) ds$, 
which implies 
\begin{equation}
  d_{31}(t) \leq \exp \int_0^t \gamma^{\prime}[a(s)] ds  
\label{vald34}
\end{equation}
by Gronwall's lemma.

Finally, 
\begin{equation}
  d_4(t) = 
  \sup_{\phi \in {\cal D}}
       \Bigl| \int_\Lambda (\phi \circ T_{t,0}[a(.)] - \phi \circ T_{t,0}[b(.)]) 
                           d \nu_0 \Bigr| \leq d_{40}(t)
\label{vald41}
\end{equation}
where 
\begin{equation}
  d_{40}(t) 
  := \sup_\Lambda \| T_{t,0}[a(.)] (x,v) - T_{t,0}[b(.)] (x,v) \|
  \leq d_{41}(t) + d_{42}(t)  
\label{d40}
\end{equation}
with 
\begin{eqnarray}
  d_{41}(t) &:=& 
  \sup_\Lambda \| \int_0^t \bigl( G[a(s)] T_{s,0}[a(.)] (x,v) -
                  G[a(s)] T_{s,0}[b(.)] (x,v) \bigr) ds \| \\
  &\leq & \int_0^t \gamma^{\prime}[a(s)] d_{40}(s) ds  
\label{d41}
\end{eqnarray}
\begin{eqnarray}
  d_{42}(t) &:=& 
  \sup_\Lambda \| \int_0^t \bigl( G[a(s)] T_{s,0}[b(.)] (x,v) -
                  G[b(s)] T_{s,0}[b(.)] (x,v) \bigr) ds \| \\
  &\leq & \int_0^t \sup_\Lambda \| G[a(s)] - G[b(s)] \| ds  
\label{d42}
\end{eqnarray}
Definition (\ref{Gxv}) shows that 
\begin{equation}
  \| G[a(s)] (x,v) - G[b(s)] (x,v) \| 
  = \bigl| {\frac{\alpha q \tau }{m}}
           \sum_j \beta^{\prime}_j (a_j(s) - b_j(s)) e^{i k_j x} \bigr| 
  \leq \tau^{-1} \| a(s) - b(s) \|  
\label{d43}
\end{equation}
Now define $\varphi (t)$, a majorant of the sum $d_1(t)+d_3(t)$, 
and $d_5(t)$, a majorant of the sum $d_2(t)+d_4(t)$ as 
\begin{equation}
  \varphi (t) = 
    \| a(0) - b(0) \| 
    + e^{\int_0^t \gamma^{\prime}[a(s)] ds}
           d_{bL} (\mu_0, \nu_0)  \label{defphi} 
\end{equation}
\begin{equation}
  d_5 (t) = d_2(t) + d_{41}(t) + d_{42}(t)  
\label{defd5}
\end{equation}
Then previous inequalities for the $d_i$ lead to 
\begin{equation}
  d_5 (t) \leq 
    \sqrt{2} \tau^{-1} \int_0^t d_{bL} (\mu_s, \nu_s) ds 
    + \int_0^t \gamma'[a(s)] d_5 (s) ds 
    + \tau^{-1} \int_0^t d_1 (s) ds
\label{d51}
\end{equation}
and $d_{bL}(\mu_s,\nu_s) + d_1(s) \leq d_1(s) + d_3(s) + d_4(s) 
                                  \leq \varphi(s) + d_5(s)$ 
so that
\begin{equation}
  d_5 (t) \leq 
    \int_0^t \sqrt{2} \tau^{-1} \varphi (s) ds + 
    \int_0^t (\sqrt{2} \tau^{-1} +
    \gamma^{\prime}[a(s)]) d_5 (s) ds  
\label{d52}
\end{equation}
which Gronwall's inequality readily estimates by 
\begin{equation}
  d_5 (t) \leq 
    \int_0^t \sqrt{2} \tau^{-1} \varphi(s) 
             e^{\int_s^t (\sqrt{2} \tau^{-1} + \gamma^{\prime}[a(u)]) du} 
             ds  
\label{vald5}
\end{equation}

The resulting complete estimate 
\begin{equation}
  \|(\mu_t, a(t)) - (\nu_t, b(t))\| \leq \varphi(t) + d_5 (t)  
\label{th1}
\end{equation}
depends on two functions $\gamma^{\prime}[a(s)]$ and $\varphi(s)$. Note
that $\varphi(0) = \|(\mu_0, a(0)) - (\nu_0, b(0))\|$ and $d_5(0) = 0$.
Estimate (\ref{th1}) does not grow faster than exponentially, with upper
bound on its growth rate 
\begin{equation}
  C = \sqrt{2} \tau^{-1} + 2 \sup_{0 \leq s \leq t} \gamma^{\prime}[a(s)]  
\label{Cestim}
\end{equation}
which is bounded by a function of $h_0$ as discussed in Section IV. This
completes the proof of the theorem.

The corollary follows in a standard way.

{\it Remark} : our estimate for the growth rate $C$ in the kinetic case is
larger than the finite-$N$ estimate for $|{\cal M}|$ in phase space. 
This is due to the fact that the distance $d_{bL}$ 
makes no distinction between $x$-components and $v$-components, 
while estimates of Sec. IV relied on treating these components 
of the phase space points separately to obtain (\ref{maxM2}).

\section{Conclusion} This work supports theoretically the use of full $N$%
-body dynamical schemes \cite
{Cary92,Cary93,Escande96,Guyomarch95,Guyomarch96a,Guyomarch96b} to study the
wave-particle interactions, as an alternative to kinetic-theory based
models. However the regularity of the limit $N \to \infty$ is tempered by
the rapid growth of the right hand side in the upper bound (\ref{expothm}).

It also identifies the fundamental cause of phase space mixing and approach
to equilibrium in this many-body system~: particles passing near the
instantaneous saddle points associated with the modes undergo exponential
dichotomy, with a divergence rate controlled by amplitudes $|z_j|=|a_j|$.
This implies that the phase space regions where discrepancies between the
kinetic description and the finite-$N$ description show up most rapidly
correspond to the neighbourhood of the `separatrices' associated with the
envelopes in the particles' $\mu$-space $\Lambda$, as was observed in
numerical simulations for $M=1$ by Guyomarc'h \cite{Guyomarch96a,Guyomarch96b}.

\section{Acknowledgments}

The authors thank F.~Doveil, D.~Fanelli, D.~Guyomarc'h and P.~Bertrand for
fruitful discussions. MCF is supported by a grant from the Minist\`ere de
l'enseignement sup\' erieur et de la recherche.


\end{document}